\documentclass[pra,twocolumn,showpacs,amsmath,amssymb]{revtex4}%
\usepackage{color}
\usepackage{graphicx}
\usepackage{amsmath}
\usepackage{amsfonts}
\usepackage{amssymb}%
\providecommand{\U}[1]{\protect\rule{.1in}{.1in}}

\begin{document}

\title{Bloch oscillations and quench dynamics of interacting bosons in an optical lattice}

\author{K. W. Mahmud,$^{1}$ L. Jiang,$^{1}$ E. Tiesinga,$^{1}$ and P. R. Johnson$^{2}$}

\affiliation{$^{1}$Joint Quantum Institute, National Institute of
Standards and Technology and University of Maryland, 100 Bureau
Drive, Mail Stop 8423, Gaithersburg, Maryland 20899, USA}
\affiliation{$^{2}$Department of Physics, American University,
Washington, DC 20016, USA}

\begin{abstract}
We study the dynamics of interacting superfluid bosons in a one
dimensional vertical optical lattice after a sudden increase of
the lattice potential depth. We show that this system can be
exploited to investigate the effects of strong interactions on
Bloch oscillations. We perform theoretical modelling of this
system, identify experimental challenges and explore a new regime
of Bloch oscillations characterized by interaction-induced
matter-wave collapse and revivals which modify the Bloch
oscillations dynamics. In addition, we study three dephasing
mechanisms: effective three-body interactions, finite value of
tunneling, and a background harmonic potential. We also find that
the center of mass motion in the presence of finite tunneling goes
through collapse and revivals, giving an example of quantum
transport where interaction-induced revivals are important. We
quantify the effects of residual harmonic trapping on the momentum
distribution dynamics and show the occurrence of
interaction-modified temporal Talbot effect. Finally, we analyze
the prospects and challenges of exploiting Bloch oscillations of
cold atoms in the strongly-interacting regime for precision
measurement of the gravitational acceleration $g$.
\end{abstract}


\maketitle

\section{Introduction}

Ultracold atoms in optical lattices can simulate many of the
phenomena associated with electrons in a periodic potential.
Compared to real crystals, however, these artificial crystals made
from laser light offer versatile control of system parameters such
as the lattice depth, geometry and particle
interactions~\cite{bloch08}. Furthermore, long coherence times,
absence of impurities, and low dissipation make them an ideal
system to observe non-equilibrium quantum
dynamics~\cite{polkovnikov11,oberthaler06}. One example is the
observation of collapse and revival dynamics of bosonic matter
wave coherence in a suddenly raised (\emph{quenched}) optical
lattice~\cite{greiner02b,will10,will11}. Another example is the
observation of Bloch oscillations, periodic motion in momentum and
real space, of ultracold atoms in an accelerating
potential~\cite{raizen96,salomon96,raizen97}. These two examples
involve two different aspects of nonequilibrium dynamics: the
single-particle physics of Bloch oscillations (BO) and the
multi-particle physics of collapse and revival (CR) coherence
oscillations which depend on atom-atom
interactions~\cite{will10,tiesinga11,mahmud13}.

Bloch oscillations arise when a constant force is applied to
particles in a periodic potential~\cite{bloch28}. They have been
observed in many physical systems, including semiconductor
superlattices~\cite{feldmann92} and ultracold
atoms~\cite{raizen96,salomon96,kasevich98,nagerl08,tino11}. Bloch
oscillations have also been used as a tool to explore band
structures and their topological
properties~\cite{zakphase12,diracmerge12}, make precision
measurements of gravity \cite{nagerl08,tino11,tino12a}, and have
been suggested as a probe to identify quantum
phases~\cite{gorshkov11,rigol13}. Although the single-particle
physics of Bloch oscillations is well understood, there are still
open questions regarding the role of particle-particle
interactions~\cite{kolovsky03,buchleitner03,corrielli13,khomeriki10}.

In this paper, we perform a theoretical study of the dynamics of
interacting ultracold bosons in a one-dimensional optical lattice
whose axis is vertically aligned with gravity. Transport in two
horizontal directions is suppressed. This system, which has been
explored in several recent
experiments~\cite{greiner02b,will10,will11}, is ideally suited for
studying the interplay between particle-particle interactions and
Bloch oscillations physics. We consider a quench scenario where,
starting from an initial superfluid state, the lattice depth is
suddenly increased so that tunneling is suppressed, the atom
density frozen, and we are in the strong field regime $F\gg J,$
where $F$ and $J$ are the gravitational potential energy
difference and tunneling energy between two neighboring lattice
sites, respectively. We show that the gravity-induced Bloch
oscillations are strongly modified by interaction-induced
matter-wave collapse and revivals.

In a deep lattice with negligible tunneling and higher-band
excitation, the dynamics involves on-site phase evolution governed
by the competing and independent effects of $F$ and $U,$ the
two-body interaction energy. We study the dynamics in two limits
-- the strong-$U$ ($U>F$) regime, and the strong-$F$ ($F>U$)
regime. Our analysis provides a unified theory for interacting BO
which treats all regimes, and makes predictions that should be
within reach of future experiments. Experiments in the strong-$F$
regime have recently been performed by F. Meinert \emph{et
al.}~\cite{nagerl13}.

We also investigate three dephasing mechanisms: (i) effective
three-body interactions, (ii) finite value of tunneling, and (iii)
residual harmonic trapping. In particular, we model in detail the
momentum and real space oscillations of a lattice-trapped
superfluid in the presence of both gravity and a background
harmonic potential. We find that the dephasing effect due to
effective three-body interactions becomes important for the
strong-$U$ regime. When $J\neq0$, we predict that the Bloch
oscillations of the center of mass of the atomic cloud should also
go through collapse and revivals, demonstrating a novel
interaction-induced effect on quantum transport. We quantify how
the presence of a harmonic trap during the dynamics quickly
destroys coherence visibility, although we show that there can
also be interaction-modified temporal Talbot
revivals~\cite{chapman95,nagerl10}.

We are also interested in the prospects for using Bloch
oscillations of cold atoms for precision measurement of $g.$ Most
experiments have previously focused on the mean-field regime
~\cite{nagerl08,tino11,tino12a}, where up to 20000 BO have been
observed, although very recently experiments~\cite{nagerl13} have
operated within the strongly-correlated, deep lattice regime. We
present estimates for the bounds on the residual harmonic trapping
and finite tunneling that should allow observations of up to 50000
Bloch oscillations.

Most previous studies of BO of ultracold atoms have used the
Gross-Pitaevskii equation to model the mean-field regime when the
number of particle per lattice site is on the order of hundreds or
thousands~\cite{oberthaler06,kasevich98,nagerl08}. In contrast, we
use the Bose-Hubbard Hamiltonian, and time-evolving block
decimation (TEBD) algorithm~\cite{vidal03} for our numerical
simulations, to model the dynamics when there are a few atoms per
lattice site and particle correlations need to be properly
accounted for. We also obtain analytical approximations in the
limits of coherent states and the Thomas-Fermi regime. Bloch
oscillations for interacting bosons in this regime have been
studied by Kolovsky and
collaborators~\cite{kolovsky03,kolovsky04,kolovsky04b,kolovsky07,kolovsky09,witthaut05},
and the transport properties of Mott insulators under a constant
force~\cite{sachdev02,greiner11,rubbo11} and superfluids in a
Galileo ramp~\cite{collura12} have also been investigated. Our
focus here is on regimes where matter-wave collapse and revivals
due to interactions is important.

The article is organized as follows. In Sec. II we present our
model, define observables, and describe our computational methods.
In Sec. III, we briefly consider collapse and revivals dynamics in
a mean-field theory when the initial state is a coherent state. In
Sec. IV, we present our results for the Bloch oscillations of
strongly-correlated interacting bosons in a vertical lattice. In
Sec. V, we investigate dephasing from effective three-body
interactions, finite tunneling and residual harmonic trapping. In
Sec. VI, we analyze the prospects for precision measurement of
gravity. Finally, we summarize our results in Sec. VII.

\section{Model and Methods}
\label{sec:initial}

\subsection{System}

\begin{figure}[ptb]
\vspace{-0.1cm}
\par
\begin{center}
\includegraphics[width=0.40\textwidth,angle=0]{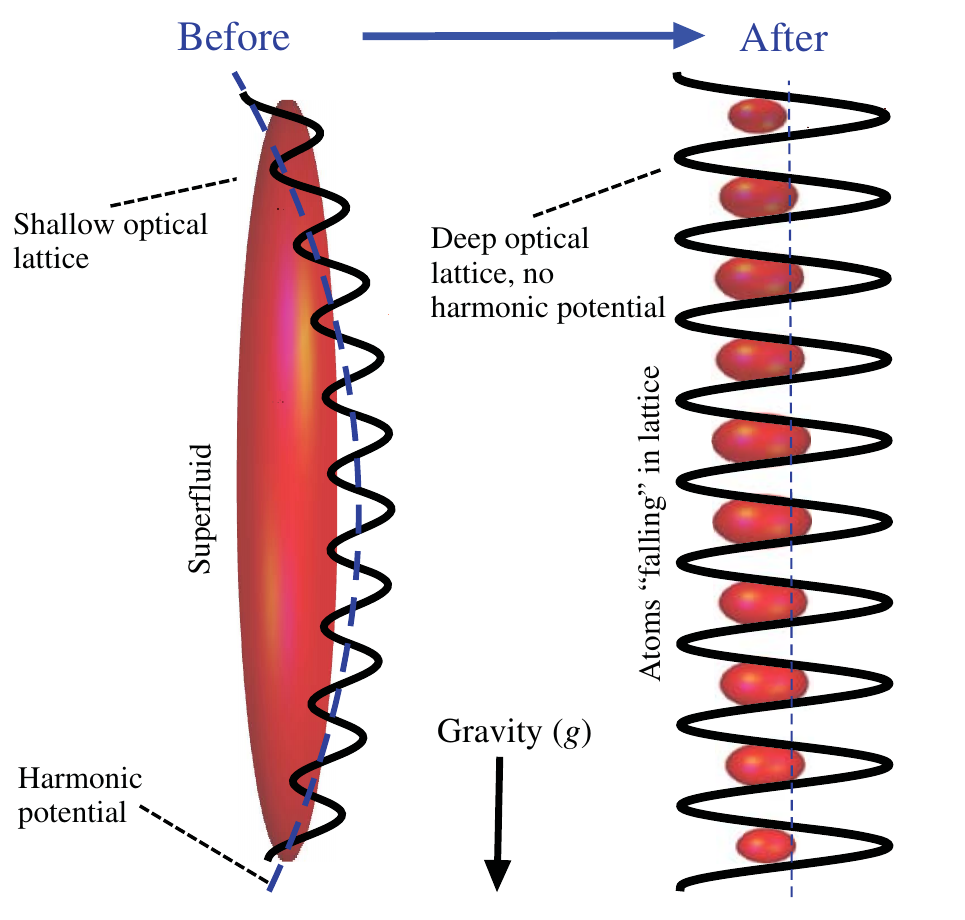}
\end{center}
\par
\vspace{-0.4cm}\caption{(color online) Schematic of system. We
start with a superfluid ground state in a shallow, vertically
aligned optical lattice. A harmonic trap supports the atoms
against gravity. The lattice depth is then suddenly increased and
simultaneously the harmonic potential is turned off. These steps
create a non-equilibrium state of atoms \textquotedblleft
falling\textquotedblright\ in the lattice under the influence of
gravity and tunable atom-atom interactions.}
\label{fig:setup}%
\end{figure}

We consider quasi-one dimensional bosons in the lowest band of a
periodic or lattice potential with period $d$. We assume that the
particles are tightly confined in the two transverse directions
such that tunneling and transverse excitations are negligible.
Under these assumptions the system is initially described by the
Bose-Hubbard Hamiltonian,
\begin{eqnarray}
H_{i} &  =&-J_{i}\sum_{j}\left(  a_{j}^{\dagger}a_{j+1}+a_{j+1}^{\dagger}%
a_{j}\right)  +\dfrac{U_{i}}{2}\sum_{j}n_{j}\left(  n_{j}-1\right)
\nonumber\\
&&  +V_{T,i}\sum_{j}j^{2}\times n_{j}-F_{i}\sum_{j}j\times n_{j},\label{HubbB}%
\end{eqnarray}
where $a_{j}^{\dagger}$, $a_{j}$ are boson creation and
annihilation operators at lattice site $j$,
$n_{j}=a_{j}^{\dagger}a_{j}$ is the boson number operator, $J_{i}$
is the initial tunneling energy (hopping parameter) between
nearest neighbors, and $U_{i}$ is the initial on-site
particle-particle interaction energy. In addition to the lattice
potential, we include an external harmonic potential initially
parameterized by energy $V_{T,i}$. The gravitational potential
energy difference between neighboring lattice sites is
$F_{i}=mgd,$ where $m$ is the atom mass, $g$ is the acceleration
of gravity, and $d=\lambda/2$ and $\lambda$ is the wavelength of
the laser that creates the periodic potential. For $^{87}$Rb and a
laser with $\lambda=$ 738 nm, the gravitational energy is
$F_{i}/h=774$ Hz, where $h$ is Planck's constant. The tunneling
energy $J_{i}$ can be tuned by changing the lattice depth
(typically $3E_{R}$ to $41E_{R},$ where
$E_{R}=\hbar^{2}k^{2}/(2m)$ is the one-photon recoil energy, and
$\hbar=h/(2\pi)$). The interaction energy $U_{i}$ depends weakly
on lattice depth, but can be tuned via a Feshbach
resonance~\cite{chin10}. A magnetic field gradient can be applied
to tune the value of $F_{i},$ as in~\cite{nagerl13}.

We start with a superfluid ground state in a shallow, vertically
aligned optical lattice (see Fig.~\ref{fig:setup}). The atoms are
initially supported against gravity by a harmonic potential. The
linear potential of gravity shifts the minimum of the harmonic
well, and there are no Bloch oscillations in the initial ground
state. The depth of the optical lattice is then suddenly increased
so that tunneling is suppressed. The parameter values before and
after the quench are labelled by subscripts $i$ and $f$,
respectively. In the ideal quench scenario, tunneling is turned
off ($J_{i}\rightarrow J_{f}=0$), and simultaneously the harmonic
trap is switched off ($V_{T,i}\rightarrow V_{T,f}=0,$ e.g., using
the methods in~\cite{will10}). The lattice ramp-up is assumed fast
compared to atom-atom interactions, yet slow enough to prevent
excitations to higher bands. The post-quench ideal final
Hamiltonian is then
\begin{equation}
H_{\rm ideal}=\dfrac{U_{f}}{2}\sum_{j}n_{j}\left(  n_{j}-1\right)
-F_{f}\sum _{j}j\times n_{j},\label{Hfinal}
\end{equation}
where $U_{f}$ and $F_{f}$ denote the interaction and gravitational
energy parameters after the quench. These steps create a
nonequilibrium state of the atoms \textquotedblleft falling
\textquotedblright in the lattice. In contrast, in
Ref.~\cite{nagerl13} the system is \textquotedblleft quenched
\textquotedblright\ by suddenly changing $F_{i}\rightarrow F_{f}$
by changing an applied magnetic field gradient. Here we have taken
$F_i=F_f=F$ throughout, since we focus on measuring $g$.

After a lattice hold time $t_{h}$, observables are measured either
in-situ~\cite{greiner09} or through time of flight
imaging~\cite{will10}. To see the effects of gravity on the atoms,
imaging has to be done from the side as opposed to from the top or
bottom.

We also model more realistic experimental conditions where there
is residual harmonic trapping and finite tunneling. To find the
ground states and simulate the time evolution, we use the
time-evolving block decimation (TEBD) algorithm~\cite{vidal03}.
This is a near-exact numerical method where we can control the
accuracy of our simulations. The TEBD algorithm is based on a
matrix product state Ansatz and is equivalent to time-dependent
density matrix renormalization group (DMRG) methods.

\subsection{Observables}

To analyze the non-equilibrium dynamics, we follow observables
giving the center of mass position, momentum distribution, zero
momentum occupation, and condensate fraction. The center of mass
position $x_{cm}$ (in units of $d$) is determined from density
measurements as
\begin{equation}
x_{cm}(t)=\frac{1}{N}\sum_{j=1}^{L}j\langle n_{j}(t)\rangle,
\end{equation}
where $N=\sum_{j} \langle n_j \rangle$ is total atom number and
$L$ is the total number of lattice sites.

The momentum distribution can be measured using time-of-flight
expansion and is given by
\begin{equation}
\langle n_{k} \rangle=\frac{1}{L}\sum_{i,j}e^{ik(i-j)}g(i,j),
\end{equation}
where $g(i,j)=\langle a_{i}^{\dagger}a_{j}\rangle$ is the
single-particle density matrix (or Green's function). As a special
case, the occupation of the zero momentum mode is given by
$\langle n_{k=0} \rangle =(1/L)\sum_{i,j}g(i,j)$. For Bloch
oscillations the momentum peak translates in $k$-space, and we
define visibility as the occupation of the peak momentum denoted
by $\langle n_{k,max} \rangle$.

Finally, we analyze the condensate fraction $f_{c},$ which is
defined as the largest eigenvalue of the single-particle density
matrix $g(i,j)$, divided by $N$. This is a measure of the presence
of Bose-Einstein condensation in an interacting many-body
system~\cite{penrose56}.

In our treatment of time dependence of observables, we normalize
the momentum distribution with its maximum value at initial time
and define, $\langle \tilde{n}_{k} \rangle=\langle n_{k}
\rangle/\langle n_{k=0} \rangle_{t=0}$. Similarly, we normalize
other observables with the corresponding initial values and define
$\langle \tilde{n}_{k,max} \rangle$, $\langle \tilde{n}_{k=0}
\rangle$ and $\tilde{f}_c$, to facilitate comparisons.

\section{Coherent State Dynamics}

We can obtain analytic expressions for the collapse and revival
dynamics if we assume that the initial superfluid is a product of
coherent states $| \alpha_j \rangle$ in each site $j$. Such a
state could be achieved experimentally if $U_{i}$ is initially
tuned to near zero. If a coherent state is suddenly projected
(quenched) into a deep optical lattice, the resulting
non-equilibrium state shows collapse and revival in coherence due
to interactions, as first observed in Ref.~\cite{greiner02b}.

The momentum distribution after the quench is then given by
\begin{equation}
\langle n_{k}(t)\rangle=\frac{1}{L}{\left|\sum_{j}\langle
a_{j}^{\dagger}(t)\rangle
e^{-ijk}\right|^{2}}-\frac{1}{L}{\sum_{j}|\langle
a_{j}^{\dagger}(t)\rangle|^{2} }+\bar{n} \label{inithomo}
\end{equation}
where $\bar{n}=N/L$ is the average atom occupation per site.

The Hamiltonian governing the post-quench dynamics when $J=0$, is
$H_f=H_{\rm ideal}+V_{T,f}\sum_{j}j^{2} n_{j}$, and the
annihilation operator in the Heisenberg picture simplifies to
\begin{equation}
a_{j}(t)=e^{iH_ft/\hbar}a_{j}e^{-iH_ft/\hbar}=e^{-i(U_f
n_{j}+V_{T,f}
 j^2-F j)t/\hbar}a_{j} \label{anni}
\end{equation}

If we assume that the lattice is homogeneous and large, then
$\langle a_{j}(t)\rangle=\alpha\exp[|\alpha|^{2}(e^{-iU_f
t/\hbar}-1)] e^{i(-Fj+V_{T,f}j^{2})t/\hbar}$. From here we can
define the quantity,
\begin{equation}
v(t)=|\langle a(t)\rangle|^{2}=\bar{n}e^{2\bar{n}\left[cos(U_f
t/\hbar)-1\right]},
\end{equation}
where $\bar{n}=|\alpha|^{2}$ and the CR oscillation period is
$T_{U}=h/U_f$.

When only the gravitational potential is present during the
dynamics, the momentum distribution for a homogeneous system is
given by,
\begin{equation}
\langle n_{k}(t)\rangle=\left(  \frac{1}{L}\frac{\sin^{2}\left[
\left( kd+\omega_B t\right)  L/2\right]  }{\sin^{2}\left[  \left(
kd+\omega_B t\right)  /2\right] }-1\right)  \times v(t)+\bar{n}.
\label{eq:decouple}
\end{equation}
where $\omega_B=F/\hbar$. A similar expression can be derived when
the initial density $n_{j} =|\alpha_{j}|^{2}$ depends on position
(e.g., for a Thomas-Fermi initial profile).

When only a harmonic trap is present during the evolution, we can
obtain analytic expressions for the dynamics of $\langle n_{k=0}
\rangle$ in two different approximations: (i) assuming a
homogenous pre-quench state, that is $n_{j}=\bar{n},$ and (ii)
assuming a Thomas-Fermi initial density profile. For case (i), we
obtain
\begin{equation}
\langle
n_{k=0}(t))\rangle=\frac{1}{L}{v(t)|\sum_{j}e^{iV_{T,f}j^{2}t/\hbar}|^{2}
}-v(t)+\bar{n}.\label{eq:inithomog}
\end{equation}
For case (ii), we use a Thomas-Fermi profile, $n_{j}=\bar{n}(1-(V_{T,i}%
/\mu_{o})j^{2})$, where $\bar{n}=\mu_{0}/U_{i}$ and $\mu_{0}$ is
the chemical potential. Taking the continuum limit the sum turns
into an integral, and after the change of variables
$y=j\sqrt{V_{T,i}/\mu_{0}}$, we obtain
\begin{align}
&  \langle n_{k=0}(t)\rangle=\bar{n}-v(t)+\frac{1}{L}{v(t)e^{2
\bar{n}}
}\times\nonumber\\
&  {\left(\frac{\mu_{0}}{V_{T,i}}\right)^D\left | \int^{b}_{-b}
y^{D-1}dy\sqrt{1-y^{2}}e^{(-\bar {n}+\frac{it\mu_{0}}{\hbar}
\frac{V_{T,f}}{V_{T,i}})y^{2}} \right |^{2}}. \label{inittrapped}
\end{align}
Here $D$ is the dimensionality of the system and
$b=\frac{L}{2}\sqrt{V_{T,i}/\mu_0}$. We use
Eqs.~\ref{eq:inithomog} and~\ref{inittrapped} in Sec.~\ref{sec:g}
to model the early time decay of the zero momentum occupation, and
analyze the effects of a residual harmonic trap on measuring $g$.

\section{Dynamics in a vertical lattice and $J_f=0$}
\label{sec:dynamics}

\begin{figure}[ptb]
\vspace{-0.1cm}
\par
\begin{center}
\includegraphics[width=0.35\textwidth,angle=0]{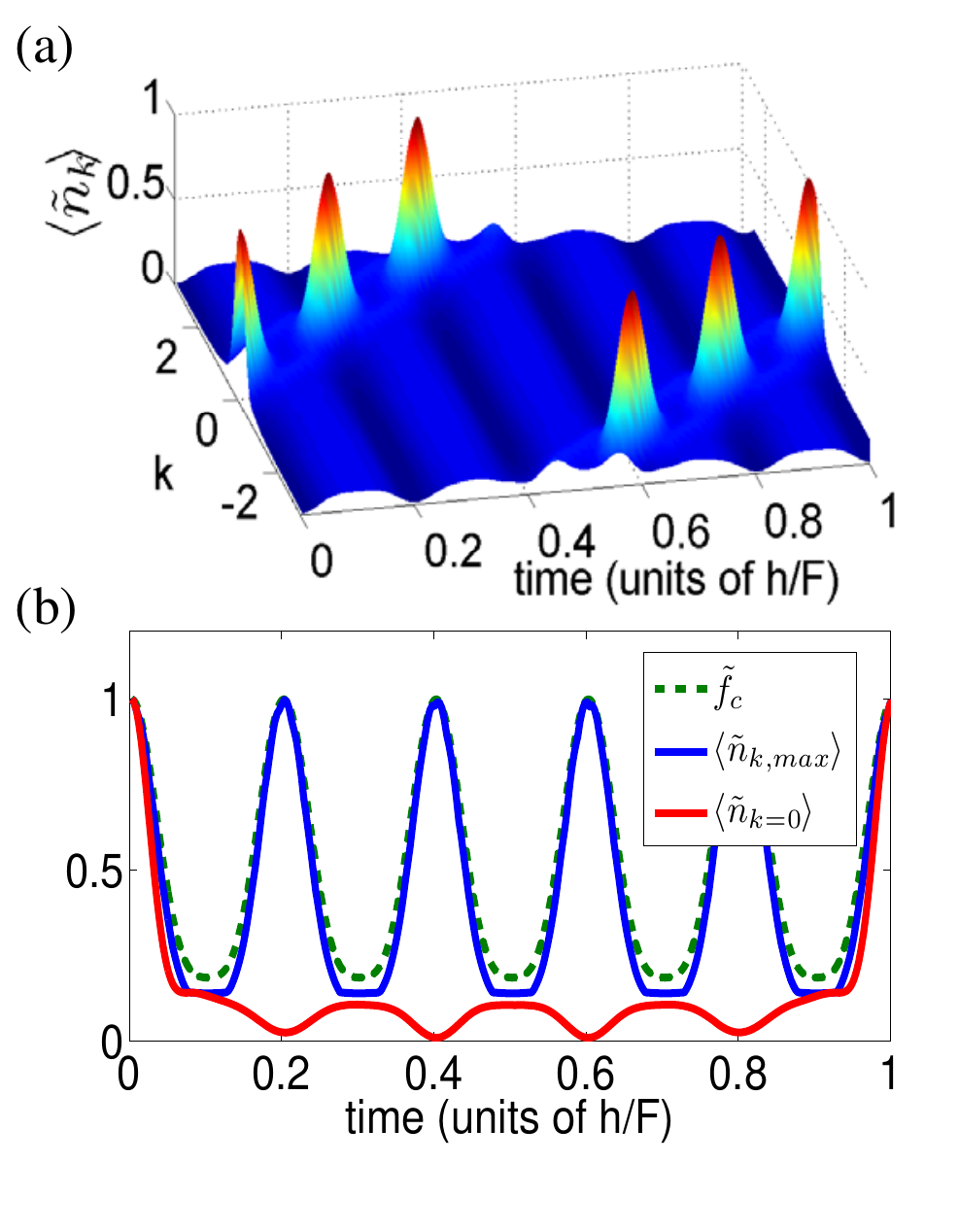}
\end{center}
\par
\vspace{-1.0cm}\caption{(color online) Bloch oscillations collapse
and revival dynamics in the strong$-U$ regime when $U_{f}>F$. Here
$U_{f}=5F$. Shown are the dynamics of (a) the quasi-momentum
distribution and (b) the peak momentum occupation and condensate
fraction, as a function of hold time $t_h$ in the lattice, given
in units of $h/F$. Panel (a) shows that the atomic momentum
performs two kinds of evolution: Bloch oscillations (BO) and
collapse and revivals (CR) of coherence. The momentum peak travels
in quasi-momentum space reaching the end of the Brillouin zone at
$k=\pi$, reflecting to $k=-\pi$ and coming back to $k=0$ to
perform one Bloch oscillation with period $h/F$. During this time
interval, the momentum peak also collapses and revives with period
$h/U_{f}.$ During collapse, atoms are distributed in
quasi-momentum over the entire Brillouin zone. The observables in
Panel (b) also reveal the simultaneous presence of BO and CR.
Dynamics of condensate fraction $\tilde{f}_c$ is not affected by
the linear potential.}
\label{fig:strongU}%
\end{figure}

In this section, we analyze the dynamics using a TEBD algorithm
under idealized conditions where, after the quench, $J_{f}=0$ (no
tunneling) and $V_{T,f}=0$ (no residual harmonic potential). We
consider both strong-$U$ ($U_{f}>F$) and strong-$F$ ($F>U_{f})$
regimes.

Figure~\ref{fig:strongU} shows the post-quench dynamics in the
strong-$U$ regime. For the initial superfluid, we choose
$U_{i}/J_{i}=3$ and lattice size $L=32$. The initial atomic cloud,
before the quench, is supported against gravity by a harmonic
potential. We choose $V_{T,i}=0.02 U_i$ and $N=40$. Unless
otherwise noted, all figures will use these pre-quench values. To
induce Bloch oscillations, the harmonic potential is turned off
simultaneously with the lattice ramp. After the quench, we set
$U_{f}=5F$, which corresponds to $F=774$ Hz and $U_{f} \approx 4$
kHz. The collapse and revival experiment of Ref.~\cite{will10}
would fall in this strong-$U$ regime.

Figure~\ref{fig:strongU}(a) shows how the quasi-momentum
distribution manifests two distinct behaviors. First, the peak or
the center of the distribution moves uniformly in momentum space
following $k(t)=k(0)+mgt/\hbar$; this results from the
gravitational acceleration $g$. When it reaches the Brillouin zone
boundary at $k=\pi$ it is Bragg scattered to $k=-\pi$. The motion
continues and the peak returns to its original position at $k=0$
in one BO period $T_{B}=h/F$. In this case of an infinitely deep
lattice (i.e., $J_{f}=0$)$,$ the atomic spatial density is frozen
and the Bloch motion appears only in momentum space. The dynamics
is driven by the relative gravitational phase shift
$e^{imgdt/\hbar}$ between each neighboring sites.

Second, Fig.~\ref{fig:strongU}(a) shows that the momentum peak
undergoes interaction-driven collapse and revival oscillations,
with revival time $T_{U}$. In this example, there are five CR
oscillations per BO, as $U_{f}/F=5.$ Here the BO and CR
oscillations are decoupled; the analytic expression in
Eq.~(\ref{eq:decouple}) expresses this concisely. If $U_{f}=0$ (no
atom-atom interactions, which could be achieved experimentally via
a Feshbach resonance), the momentum peak traverses the Brillouin
zone with no CR oscillations. The role of interactions in BO in
causing collapse and revivals is the same as the role interactions
play in the CR experiments~\cite{greiner02b,will10}.

CR oscillations can also be seen in a measurement of visibility by
monitoring the momentum peak evolution $\langle \tilde{n}_{k,max}
\rangle$ as depicted in Fig.~\ref{fig:strongU} (b). We also plot
the condensate fraction $\tilde{f}_{c}$ and $\langle
\tilde{n}_{k=0} \rangle$. The condensate fraction and visibility
are closely related and proportional~\cite{hofstetter11}. The
condensate fraction dynamics shows that the interaction-induced
quantum depletion is decoupled from the evolution generated by the
linear potential due to gravity or a uniform applied magnetic
field.

\begin{figure}[ptb]
\vspace{-0.1cm}
\par
\begin{center}
\includegraphics[width=0.35\textwidth,angle=0]{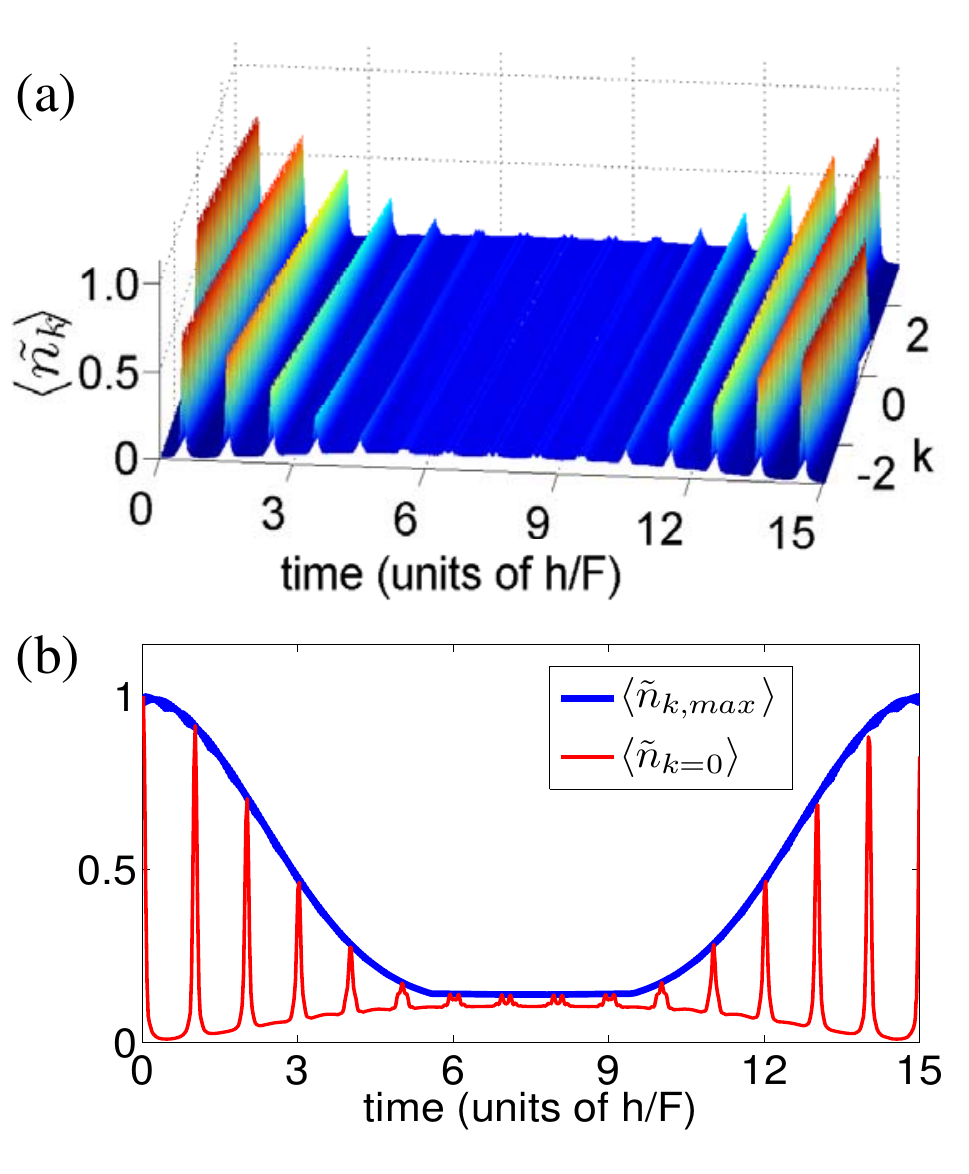}
\end{center}
\par
\vspace{-0.8cm}\caption{(color online) Bloch oscillations collapse
and revival dynamics in the strong-$F$ regime when $F>U_{f}$. Here
$F=15U_{f}$. Panel (a) shows the quasi-momentum population
$\langle \tilde{n}_k \rangle$ versus hold time. For this example,
there is one CR cycle for every 15 BO. In this regime, $\langle
n_{k=0} \rangle$ evolution, shown in Panel (b) reveals both CR and
BO dynamics. The peak momentum evolution $\langle
\tilde{n}_{k,max} \rangle$ shows only CR.}
\label{fig:strongF}%
\end{figure}
Figure~\ref{fig:strongF} shows post-quench dynamics in the
strong-$F$ regime, which could be achieved by using a Feshbach
resonance to tune atom-atom interactions toward zero. The recent
experiment~\cite{nagerl13} was performed in this regime. The
initial superfluid corresponds to $U_{i}/J_{i}=3.$ After the
quench we set $J_{f}=0$ and $F/U_{f}=15.$
Figure~\ref{fig:strongF}(a) shows $15$ Bloch oscillations in
momentum space for every CR oscillation. Fig.~\ref{fig:strongF}(b)
shows the dynamics of both the peak momentum occupation and zero
momentum occupation, versus hold time. In this regime, collapse
and revivals occur over many BO cycles. $\langle \tilde{n}_{k=0}
\rangle$ evolution here reveals both CR and BO dynamics. CR can be
viewed here as interaction-induced dephasing, and subsequent
re-phasing, of the Bloch oscillations (see also
~\cite{kolovsky03}).

\section{Dephasing mechanisms}

\subsection{Effective 3-body interactions}

\begin{figure}[ptb]
\vspace{-0.1cm}
\par
\begin{center}
\includegraphics[width=0.47\textwidth,angle=0]{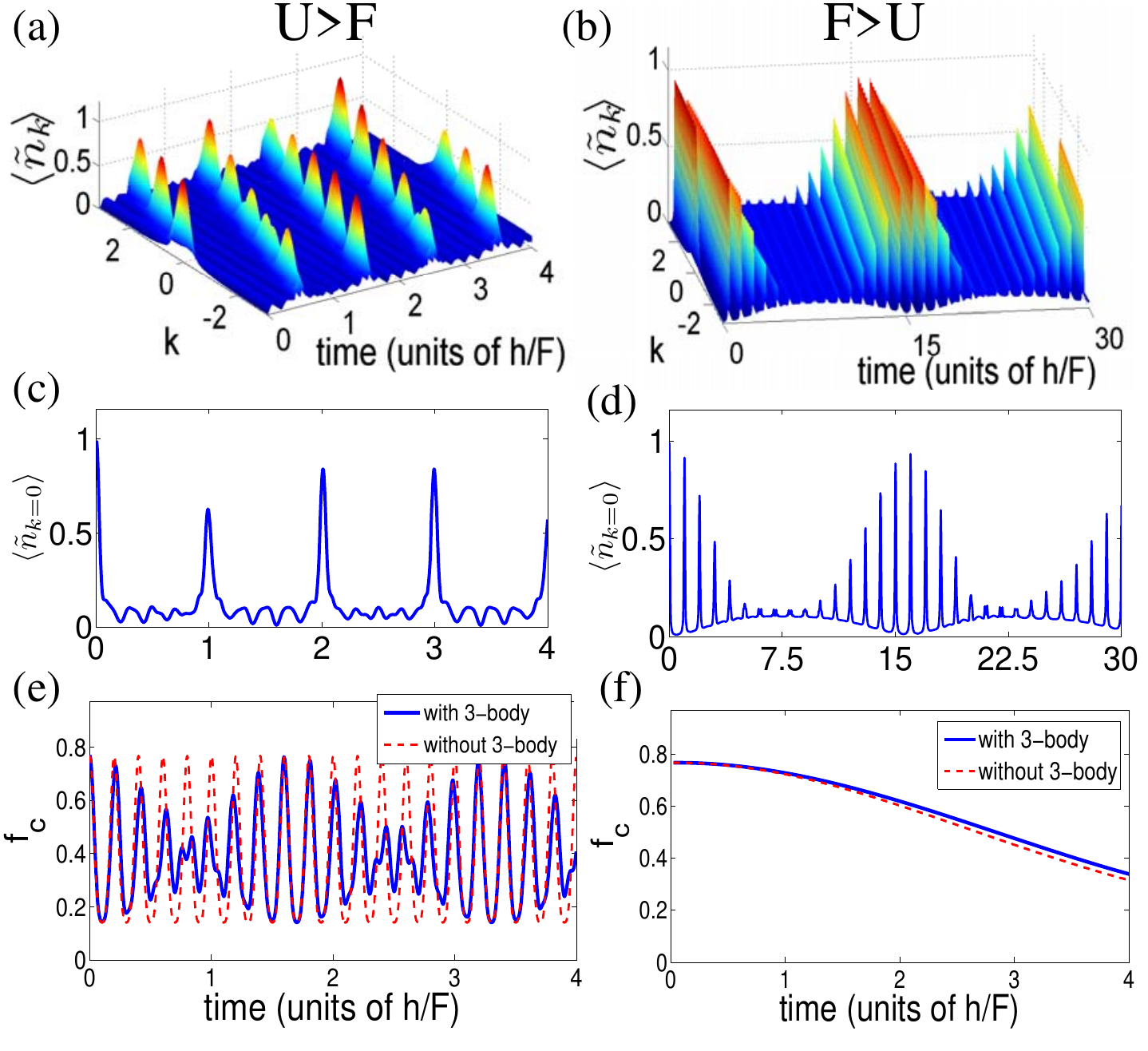}
\end{center}
\par
\vspace{-0.8cm}\caption{(color online) Influence of effective
3-body interactions on the collapse and revivals of Bloch
oscillation. Left column shows the strong-$U$ limit of
Fig.~\ref{fig:strongU}, and the right column shows the strong-$F$
limit of Fig.~\ref{fig:strongF}. From top to bottom the panels
show respectively $\langle \tilde{n}_k \rangle$, $\langle
\tilde{n}_{k=0} \rangle$ and $f_c$. The three-body interaction
strength is $U_{3}=-0.12U_{f}$. The momentum distribution and BO
dynamics is significantly modified during a single Bloch cycle for
the strong-$U$ regime. In contrast, the effect is far more gradual
for the strong-$F$ limit. This is highlighted in panels (e) and
(f).}
\label{fig:3body}%
\end{figure}

In a deep lattice there are effective multi-body interactions due
to collision induced virtual excitations to higher
bands~\cite{johnson09,johnson12,mahmud13}. Quantum phase revival
spectroscopy, based on the collapse and revival phenomenon, has
been used to detect the presence of effective higher-body
interactions~\cite{will10,will11}. Here we examine the influence
of effective three-body interactions on the Bloch oscillations CR
dynamics. To model this physics, we add to the Hamiltonian in
Eq.~(\ref{Hfinal}) the effective three-body term
\begin{equation}
H_{3B}=\frac{1}{3!}U_{3}\sum_{j}n_{j}\left(  n_{j}-1\right)
\left(
n_{j}-2\right)  \label{3body}%
\end{equation}
where $U_{3}$ is the effective 3-body interaction energy.

Figure~\ref{fig:3body} shows the effects of three-body
interactions in the strong-$U$ (left column) and strong-$F$ (right
column) regimes for $U_{f}=5F$ and $F=15U_{f}$, respectively. In
both cases we set $U_{i}/J_{i}=3$ so that we consider the same
initial state as in the previous section. Post-quench, we have
$U_{3}=-0.12U_{f}$~\cite{johnson09}. In Fig.~\ref{fig:3body} we
plot the dynamics of three observables: the quasi-momentum
distribution in the first Brillouin zone, the zero momentum
occupation, and the condensate fraction (which is proportional to
the visibility).

For the strong-$U$ case in Fig.~\ref{fig:3body} (a) and (c), we
see that the revival of the momentum peak after each BO cycle is
incomplete, due to the presence of effective 3-body interactions.
In contrast, in the strong-$F$ regime shown in the right column of
Fig.~\ref{fig:3body}, the 3-body interactions lead to only a small
modification of the oscillations, over the time interval shown. We
quantify this in panels (e) and (f) by comparing $f_c$ signals
over the same time interval, with and without three-body
interactions. The longer-period envelope in Panel (e) is due to
effective three-body interactions and shows their significant
influence in the strong-$U$ regime. In Panel (f) the modification
in the signal due to effective three-body interactions is minimal
on the same timescale. As expected, the influence of effective
three-body interactions is far more prominent for the strong-$U$
case. We note that the dephasing due to effective three-body
interactions will also show revivals unless that timescale is
longer than other dephasing mechanisms~\cite{will10,johnson12}.

\subsection{Finite Tunneling}

\begin{figure*}[ht]
\vspace{-0.1cm}
\par
\begin{center}
\includegraphics[width=0.9\textwidth,angle=0]{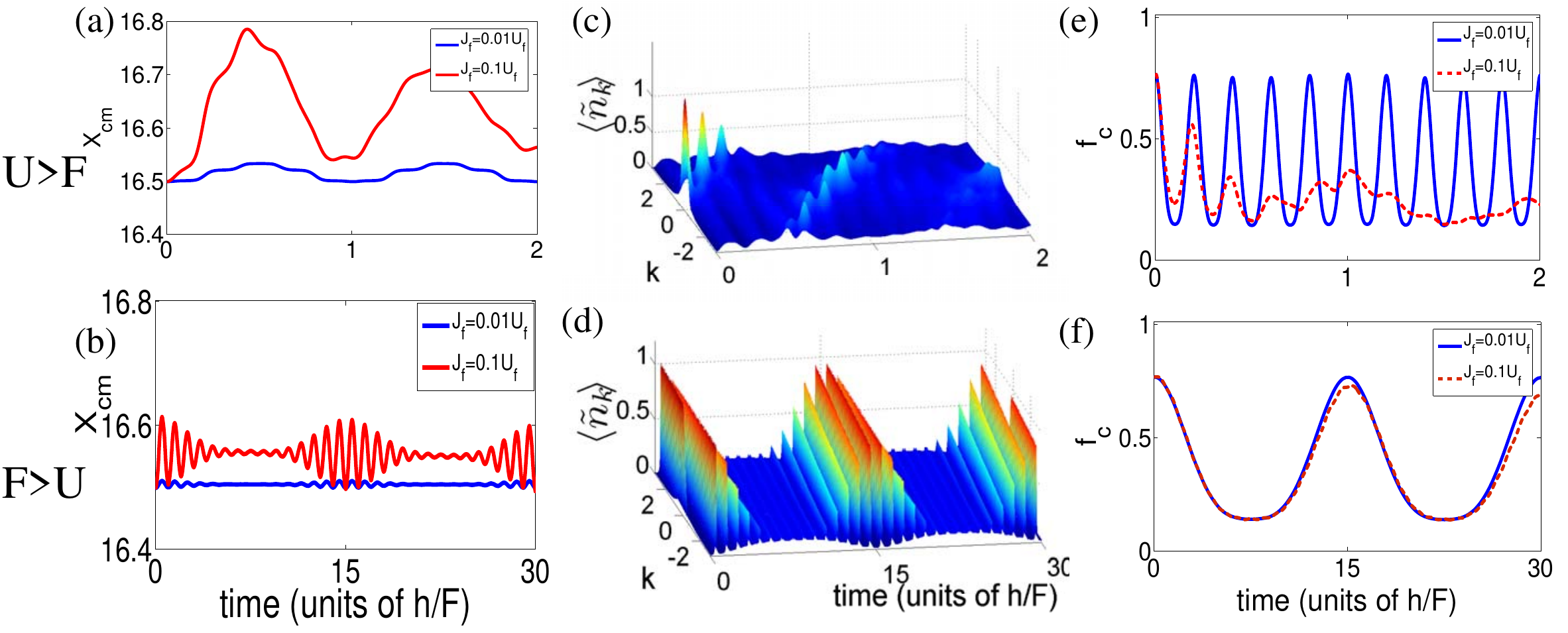}
\end{center}
\par
\vspace{-0.8cm}\caption{(color online) Effect of finite tunneling
on Bloch Oscillations revivals. Top and bottom rows show the
strong-$U$ and strong-$F$ cases respectively. The center of mass
(COM) oscillations shown in panels (a) and (b) have larger
amplitudes for larger $J_{f}$, but they also damp at a faster
rate. In Panel (b), we see that the COM performs CR oscillations,
giving an example of quantum transport where real space collapse
and revivals occur. Dynamics of momentum distribution for $J_f=0.1
U_f$ in panels (c) and (d), and condensate fraction in panels (e)
and (f) show that higher values of $J_f$ cause stronger decay in
their signals. The competition among interactions ($U_{f}$),
tunneling ($J_{f}$), and the linear potential ($F$) controls the
complex dynamics.} \label{fig:finiteJ}
\end{figure*}

When there is finite tunneling $J_f \neq 0$ between lattice sites
after the quench, Bloch oscillations manifest as position space
oscillations as well as the momentum oscillations. We consider in
this subsection finite-$J_{f}$ still assuming the regime of $F\gg
J_{f}$ and $V_{T,f}=0.$ Here

\begin{equation}
H_J=H_{\rm ideal}-J_f \sum_{j}\left(
a_{j}^{\dagger}a_{j+1}+a_{j+1}^{\dagger} a_{j}\right).
\end{equation}

Figure~\ref{fig:finiteJ} shows the effects of finite tunneling for
both strong-$U$ ($U_{f}/F=5$ in top row) and strong-$F$ regimes
($F/U_{f}=15$ in bottom row). Panels (a) and (b) show the
center-of-mass (COM) oscillations of the atomic density. Red and
blue curves show cases with larger ($J_{f}=0.1U_{f}$) and smaller
tunneling ($J_{f} =0.01U_{f}$), respectively. The influence of
interactions can be seen in the COM motion. In
Fig.~\ref{fig:finiteJ}(a), five small amplitude kinks are visible
for every BO cycle, consistent with the parameter choice
$U_{f}/F=5$. In Fig.~\ref{fig:finiteJ}(b), the collapse and
revival modulation occurs over 15 BO cycles, consistent with
$F/U_{f}=15$. We see in these simulations an example of
interaction-induced collapse and revivals for real space quantum
transport. We note that the spatial amplitude is less than a
lattice spacing, for the parameter regimes explored here. The
amplitude of the spatial oscillations is proportional to $J_{f},$
and depends on the competition of $U_{f}$ and $F$.

Figures~\ref{fig:finiteJ}(c) and (d) show the BO dynamics of the
quasi-momentum distribution for $J_{f}=0.1U_f$. In the strong-$U$
regime in Panel~(c) we observe a rapid decay of the momentum peak
caused by atoms tunneling to and interacting with atoms in
neighboring sites~\cite{daley11}. Figures~\ref{fig:finiteJ}(e) and
(f) show the dynamics of the condensate fraction $f_{\text{c}}$.
In Panel~(e), we see that the larger $J_{f}$ value leads to the
fastest damping of the condensate. For $J_f=0.1 U_f$ the BO
signals in (c) and CR signals in (e) decay significantly within
two BO periods. In contrast, the blue curve for small tunneling
shows the expected interaction-driven CR oscillations, without
significant decay of the revivals. For the strong-$F$ regime in
panels (d) and (f), we see that the decay is much slower over the
same time span.

These simulations highlight how a combination of tunneling
$(J_{f})$ and interactions $(U_{f})$ generates true damping.
Finite-$J_f$ allows tunneling to the neighboring sites and $U_{f}$
causes interactions with atoms from neighboring sites which
changes inter-site phase relationships, thus causing the overall
decay of oscillations. In single-particle BO physics with
$U_{f}=0$ and $F \gg J$, there is no damping. Similarly, in
interacting BO physics with $J_{f}=0$, there is no true damping as
the oscillations revive on the two-body timescale $U_{f}$ (and
three-body time-scale $U_{3}$). It is the combination of $J_{f}$
and $U_{f}$ in the presence of $F$ which causes dephasing. The
presence of all three energy scales ($J_f, U_f, F$) causes the
equally spaced Wannier-Stark ladders to split into a chaotic
energy spectrum with multitude of avoided crossings. This has been
identified in Refs.~\cite{kolovsky03,kolovsky04,buchleitner03} as
a reason for interaction-induced decoherence. The experiment
reported in Ref.~\cite{nagerl13} explores this phenomenon.

\subsection{Residual Harmonic Confinement}

In the system described so far, the atoms are initially supported
against gravity by a harmonic potential. To induce Bloch
oscillations after the quench, we have assumed that the harmonic
potential is turned off simultaneously with the lattice ramp,
allowing the atoms to \textquotedblleft fall\textquotedblright\ in
the lattice. Alternatively, a sudden change in applied magnetic
field (such that $F_{f}\neq F_{i}$) can be used to shift the
location of potential minimum, as in ~\cite{nagerl13}. The latter
can be done with or without a change in the harmonic confinement.
In either approach, in practice, there can remain a residual
harmonic background $V_{T,f}\neq0.$

In the CR experiments of Ref.~\cite{will10}, the harmonic
potential was minimized using a combination of red and
blue-detuned light. In Ref.~\cite{hofstetter11}, a theoretical
analysis of CR in a harmonic trap was performed with an
inhomogeneous Gutzwiller Ansatz formalism, showing rapid dephasing
for a strong harmonic background. In the context of observing the
Talbot effect with cold atoms~\cite{nagerl08,nagerl10}, harmonic
confinement is a necessary ingredient. These experiments were
analyzed using the Gross-Pitaevskii or discrete nonlinear
Schrodinger equation (DNLSE) formalisms appropriate for the
mean-field regime with many atoms~\cite{nagerl08,nagerl10}. In
this Section, we analyze, using the TEBD method, the effects of
harmonic trapping both with and without a linear force in the
strongly correlated regime.

\begin{figure}[ptb]
\vspace{-0.1cm}
\par
\begin{center}
\includegraphics[width=0.35\textwidth,angle=0]{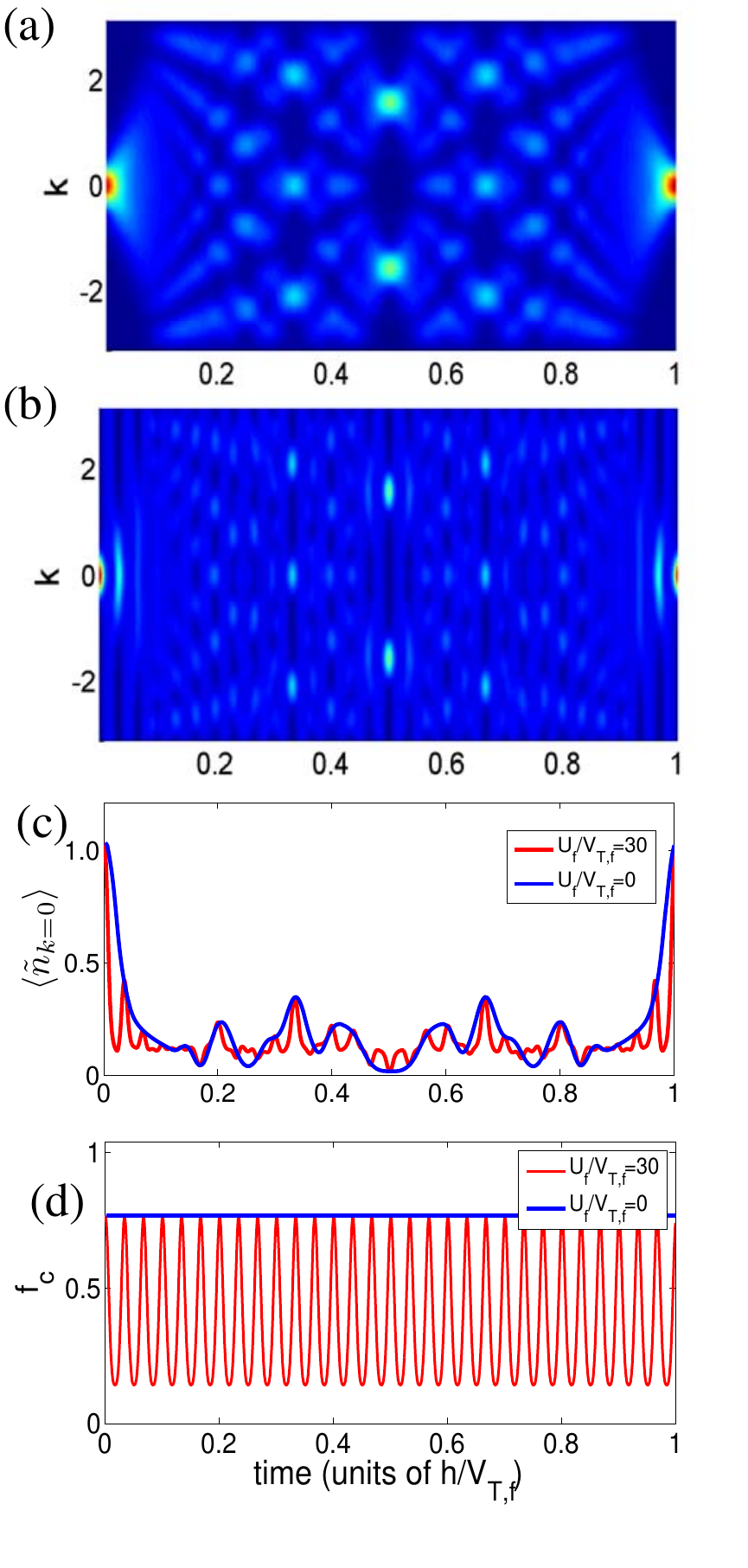}
\end{center}
\par
\vspace{-1.2cm}\caption{(color online) Collapse and revivals
dynamics in the presence of a harmonic trap. The quadratic term in
the Hamiltonian due to the harmonic potential gives rise to a
temporal Talbot effect familiar in optics. Panel (a) depicts a
density plot of $\langle n_{k} \rangle$ for noninteracting
($U_{f}=0$) system showing fractional momentum revivals. Panel (b)
shows density plot of $\langle n_{k} \rangle$ for an interacting
system where $U_{f}/V_{T,f}=30$. The interactions destroy the
Talbot revivals except at times that are integer multiples of
$h/U_{f}$, i.e., at $t=2,3,5,6,10$ in units of $h/U_{f},$ and at
symmetric times around the midpoint. Lighter colors denote higher
peaks in the momentum distribution while darker shades denote
smaller populations. Panel (c) overlays the zero-momentum
population for the above two cases. Panel (d) shows the condensate
fraction $f_{c}$ influenced only by the interactions. For
$U_{f}=0$, $f_c$ is constant.}
\label{fig:harmonicU0}%
\end{figure}

\subsubsection{Role of residual confinement without linear potential}

First, we consider the effects of only residual harmonic
confinement. This scenario can be achieved by suddenly quenching
the superfluid without reducing the harmonic background. In
Fig.~\ref{fig:harmonicU0}, we show the dynamics assuming an
initial state with $U_{i}/J_{i}=3.$ To differentiate the effects
of the harmonic trap from interactions, we show in
Figure~\ref{fig:harmonicU0}(a) a density plot of the
quasi-momentum distribution $\langle n_k \rangle$ setting
$U_{f}=0.$ The figure shows initial dephasing from the harmonic
confinement, with re-phasing (a full revival) after a period of
$T_{V}=h/V_{T,f}.$ There is also re-phasing in other quasi-momenta
at intermediate times, which give an intricate, ordered structure
called a quantum carpet~\cite{talbot11,kolovsky10}. Partial
(fractional) revivals with two, three and integer $n$ momentum
peaks are seen at $T_{V}/n$, and there are further revivals
symmetrically placed after $T_{V}/2$. The physics is analogous to
the Talbot effect~\cite{chapman95} familiar in optics, in which a
coherent state experiencing multi-site diffraction gives rise to
self-similar patterns in the near-field regime. The collapse and
revivals in Fig.~\ref{fig:harmonicU0}(a) have nothing to do with
interactions, as $U_{f}=0,$ but are due to the quadratic phase
relationship ($e^{iV_{T}j^{2}t/\hbar}$) between the neighboring
wells. Interestingly, the condensate fraction dynamics in
Fig.~\ref{fig:harmonicU0}(d) (blue line) shows that there is
always a macroscopic occupation of a single quantum state,
although the quasi-momentum has a fractal nature.

Figure~\ref{fig:harmonicU0}(b) shows density plot of $\langle
n_{k} \rangle$ when interaction is non-zero and stronger than the
harmonic confinement energy scale ($U_{f}/V_{T,f}=30$). This
figure shows the combined effects of both the harmonic potential
(with period $h/V_{T,f}$) and CR oscillations (with period
$h/U_{f})$. For the parameter choice $U_{f}/V_{T,f}=30$ there are
30 CR oscillations per harmonic period, as we can see in the
condensate fraction dynamics in Panel (d). This plot also shows
that condensate fraction dynamics is not affected by the external
harmonic potential. This physics has been explained in
Ref.~\cite{hofstetter11}: the single particle density matrix of an
inhomogeneous system is given by a unitary transformation of a
homogeneous system, and consequently the eigenvalue time-evolution
is the same in either a uniform or trapped system. The overall
quantum-carpet pattern of quasi-momentum dynamics in Panel (a) is
also seen in the strongly interacting case in Panel (b). However,
the partial or fractional Talbot revivals that persist for the
interacting case must be located at times when the
interaction-revivals also occur; in this example that happens at
factors of 30, i.e., at $t=2,3,5,6,10$ in units of $h/U_{f},$ and
at symmetric times around the midpoint.

Additional insight can be obtained through analysis of the
dynamics of the zero-momentum occupation shown in
Fig.~\ref{fig:harmonicU0}(c). The $k=0$ population quickly decays
and then revives after period $h/V_{T,f}$. The red curve
($U_{f}=30V_{T,f}$) and the blue curve ($U_{f}=0$) show that decay
of the population, driven by the harmonic background, occurs
irrespective of the value of $U_{f}$. Analyzing the early time
dependence of the population decay yields information on the
number of CR or BO cycles that can be readily observed in an
experiment. In this example, the decay is so fast that only 3 CR
oscillations can take place before harmonic dephasing dominates.

Finally, we note that there can be two other types of initial
spatial shifts that have been neglected in our simulations. The
first, trap shift, is due to the displacement of the center of the
harmonic trap within a single lattice spacing. The second, cloud
shift, is the displacement of the center of the atomic cloud from
the center of the trap caused by gravitational sag. Trap shift has
been found to influence the dynamics~\cite{hofstetter11}
introducing a linear shift in time in the momentum position in the
first Brillouin zone. These effects can be easily scaled away.

\begin{figure}[ptb]
\vspace{-0.1cm}
\par
\begin{center}
\includegraphics[width=0.36\textwidth,angle=0]{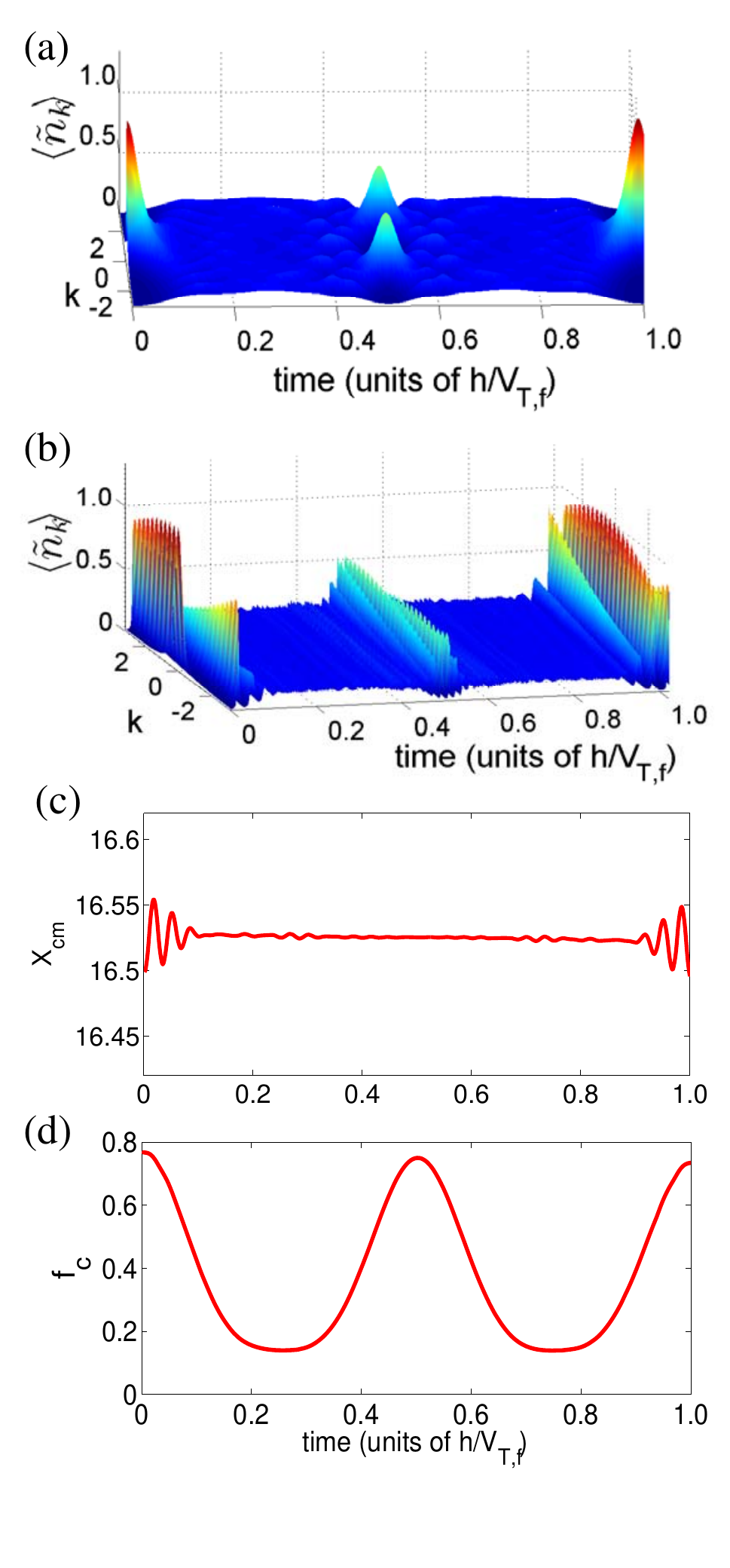}
\end{center}
\par
\vspace{-1.5cm}\caption{(color online) Effect of harmonic trapping
on Bloch oscillations revivals. The figure shows momentum
distribution dynamics without [Panel (a)] and with [Panel (b)] the
influence of a linear (e.g. gravitational) potential,
respectively, and with $F=15U_{f}$, $U_{f}=2V_{T,f},$ and
$J_{f}=U_f/4$. In Panel (b) the Talbot peaks seen in Panel (a) are
moving in $k$-space due to Bloch oscillations. Panel (c) shows the
center of mass motion, which contains signatures of interaction,
the linear accelerating potential, the harmonic trap, and finite
tunneling. Its Bloch oscillations goes through a CR sequence which
is suppressed by harmonic trap Talbot revivals at
$\frac{1}{2}h/V_{T,f}$. Panel (d) shows the condensate fraction
$f_{c}$.}
\label{fig:harmonicBloch}%
\end{figure}

\subsubsection{Role of residual confinement including gravitation (or linear)
potential}

We now analyze the dynamics when the harmonic potential is only
partially turned off during the quench, and there is a linear
external potential present such that the location of the trap
minimum suddenly shifts with the quench. We also include small but
finite tunneling ($J_{f}\neq0$). We expect the dynamics to
simultaneously manifest gravity-driven BO, interaction-driven CR
oscillations, a harmonic-background-induced Talbot effect, and the
effects of tunneling.

Figure~\ref{fig:harmonicBloch}(a) shows the quasi-momentum
distribution versus hold time when $F=0$ and $U_{f}=2V_{T,f}$. We
see the competing effects of the harmonic potential and
interactions as described earlier, with the occurrence of
fractional revivals. Figures~\ref{fig:harmonicBloch}(b), (c), and
(d) show the dynamics versus hold time with $U_{f}=2V_{T,f}$,
$F=60J_{f}$, and $F=15U_{f}$ (the strong-$F$ regime). The revivals
in Panel (b) are strongly modified by the Bloch oscillations,
which cause the momentum peaks to translate uniformly in $k$-space
and reflect at the edge of the Brillouin zone.
Figures~\ref{fig:harmonicBloch}(c) and (d) show dynamics of the
center-of-mass and condensate fraction, respectively. The real
space oscillations in Panel (c) contains signatures of all the
competing terms -- the fast modulations are due to BO, the
collapse and revival of BO is due to interaction, and the
suppression of interaction-induced revival at $\frac
{1}{2}h/V_{T,f}$ is due to harmonic trap effects. The COM motion
itself is due to finite tunneling, while the condensate fraction
($f_{c}$) dynamics in Panel (d) shows that CR oscillations are
solely due to non-zero $U_{f},$ with slow decay due to
tunneling-induced dephasing, but (again) no dependence on the
harmonic confinement. Its dephasing is also unaffected by gravity.
In our treatment of BO in this paper, the example shown here may
be most relevant to an actual experimental system since all of
these effects will be present in practice, to some degree.

\section{Measurement of $g$}
\label{sec:g}

Atomic Bloch oscillations have yielded a new method for making
precision measurements of forces. Gravitational acceleration $g$
has been measured with different degrees of precision with atomic
BECs and thermal atoms in a vertical optical
lattice~\cite{kasevich98,nagerl08,tino11,tino12a,tino12b}.
Different aspects of Bloch oscillations physics have been used for
attaining high precisions -- for example, Ref.~\cite{tino11} used
lattice modulation at the $5$th harmonic of the Bloch frequency to
induce tunneling, and Ref.~\cite{nagerl08} used a Feshbach
resonance to turn off interactions to reduce interaction-induced
dephasing due to mean-field nonlinearity. We investigate here the
prospects and challenges for the precision measurement of $g$
within a system of strongly interacting bosons in a suddenly
quenched vertical optical lattice.

The precision of the measurement of $g$ in a Bloch Oscillations
experiment depends on the number of BO cycles that can be
observed. Another factor is the narrowness of the momentum
distribution for the initial and the time-evolved state. In the
experiments of~\cite{tino11,nagerl08}, performed in the GP regime,
they were able to follow the dynamics for 10 s to 20 s and observe
approximately $20000$ BO cycles. On the other hand, experiments on
CR~\cite{will10} followed the dynamics for 20 ms, long enough time
for 20 to 30 BO cycles. Reasons for the fast decay of signals in
the CR experiments could include: pumping energy into the system
due to the quench, presence of a residual harmonic trap, presence
of finite tunneling, three-body loss and other interaction related
losses, and the effect of changing of the Wannier
function~\cite{zakrzewski13}. Here we analyze the effects of a
harmonic trap and finite tunneling, and calculate bounds on their
values for observing up to 50000 BO cycles.

\begin{figure}[ptb]
\vspace{-0.1cm}
\par
\begin{center}
\includegraphics[width=0.35\textwidth,angle=0]{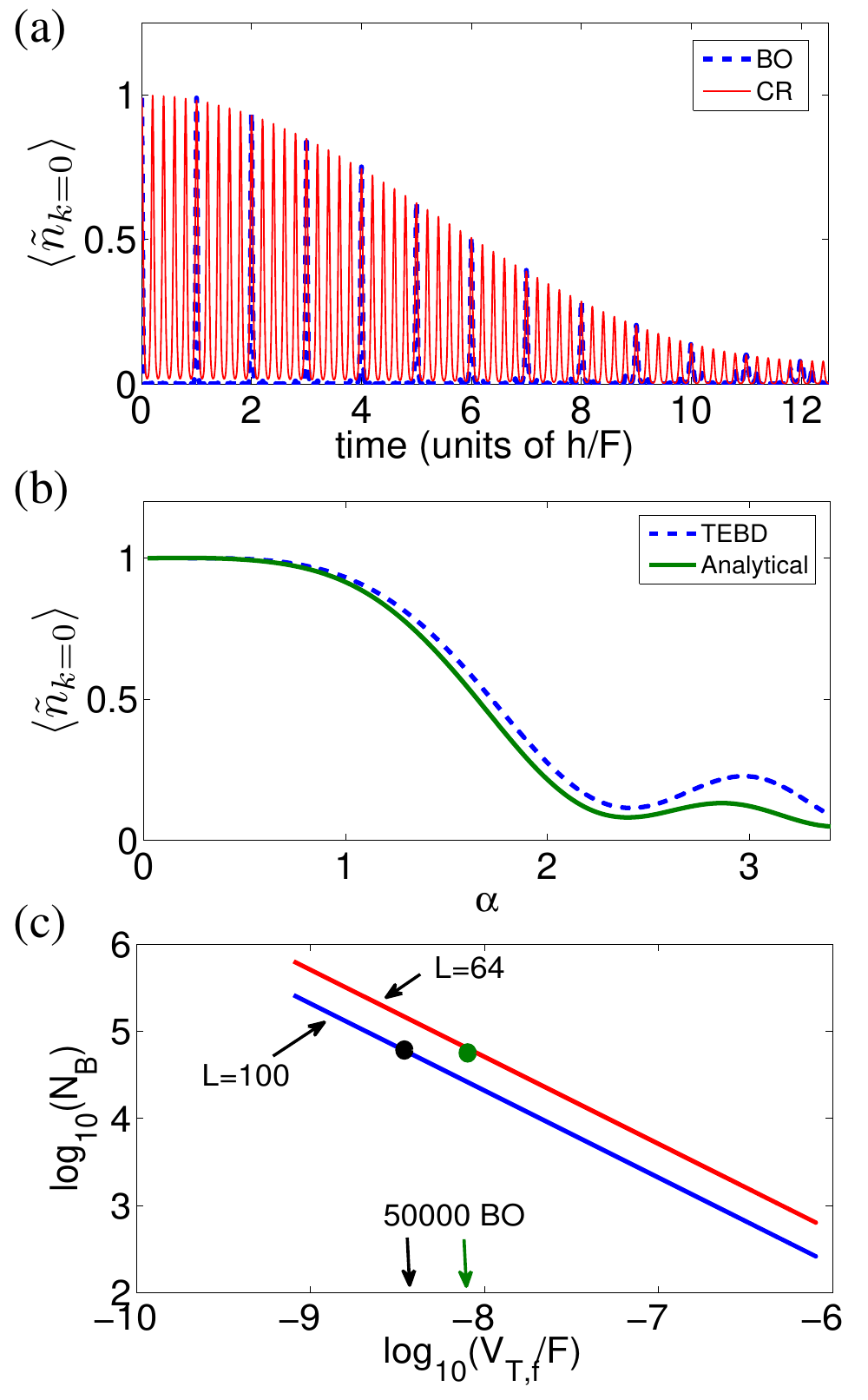}
\end{center}
\par
\vspace{-0.7cm}\caption{(color online) Bounds on the number of
Bloch oscillations due to a residual harmonic trap. Panel (a)
depicts a TEBD simulation with $L=32$ of the effects of a trap for
$V_{T,f}=0.002F$, $F=2$, and $U_f=10$. We see that both the Bloch
(dotted line) and CR (full line) oscillations with period $h/F$
and $h/U_f$ are modified by a decay envelope characteristic of the
trap strength. Here we show only the short time dynamics where the
trap-induced decay of visibility takes place. In Panel (b), we
show the analytical envelope of $\langle \tilde{n}_{k=0} \rangle$,
Eq.~(\ref{eq:errorFunc}), assuming that the initial superfluid
state is a coherent state and homogeneous, as a function of scaled
time $\alpha=L \sqrt{V_{T,f} t_h}/(2\hbar)$. We compare this to
TEBD results with $L=32$ showing a good match for the initial
decay. Panel (c) shows a $log$-$log$ plot of the number of BO
cycles when $\langle \tilde{n}_{k=0} \rangle$ drops to $1/e$ of
its initial value, as a function of residual harmonic trap
strengths for lattice sizes $L=64$ (red curve) and $100$ (blue
curve). The dots and arrows on the $V_{T,f}/F$-axis denote the
corresponding bound for 50000 BO.} \label{fig:harmonicError}
\end{figure}

\subsection{Bounds on residual harmonic trap}

We have shown in Sec. V that the presence of a residual harmonic
trap during the dynamics causes rapid decay in $\langle n_{k}
\rangle$. In separate experiments involving BO~\cite{nagerl10} and
CR~\cite{will10}, harmonic trap effects were minimized; however, a
number was not given on how small the value is. Even a minute trap
strength can have a significant effect over many oscillations.
Figure~\ref{fig:harmonicError}(a) shows a TEBD simulation of the
early time decay of $\langle \tilde{n}_{k=0} \rangle$ with
$V_{T,f}/F=0.002$, for $F=2$, $U_f=10$ and $L=32$, when the
initial state is homogeneous $V_{T,i}=0$ and final tunneling is
suppressed $J_f=0$. We see that Bloch and CR oscillations occur on
timescales $h/F$ and $h/U_f$ respectively, and they decay due to
the residual harmonic trap, following a common envelope function.
The envelope function can be understood from the approximate
continuum form of Eq.~(\ref{eq:inithomog}) assuming an initial
superfluid that is homogeneous and coherent, evolving in a
harmonic trap of strength $V_{T,f}$. The effects of $V_{T,f}$,
$U_f$ and $F$ are separable and the initial decay of the envelope
function, is given by the analytical expression
\begin{equation}
\langle n^{env}_{k=0}(t_h)\rangle \approx \pi \frac{|{\rm erf}(i
e^{i \pi/4} \alpha)|^2}{4 \alpha^2} \label{eq:errorFunc}
\end{equation}
where ${\rm erf}$ is the error function, $\alpha=L\sqrt{V_{T,f}
t_h}/(2\hbar)$, $L$ is the number of lattice sites and $t_h$ is
the hold time. Figure~\ref{fig:harmonicError}(b) shows $\langle
n^{env}_{k=0} \rangle$ as a function of the dimensionless variable
$\alpha$. The expression is universal for any harmonic trap and
lattice size; decay for different trap strengths can be scaled to
fall on this same curve. A comparison of analytical result with
TEBD simulations, which include correlations due to tunneling in
the initial superfluid, shows a good match for the initial decay.

Using Eq.~(\ref{eq:errorFunc}) we find that we can quantify the
number of BO, $N_B$, that can be observed by the hold time when
the envelope of $\langle n_{k=0}(t)\rangle$ decays to $1/e$ of its
initial value. This gives a relationship between the background
trap strength ($V_{T,f}/F$) and $N_B$,
\begin{equation}
N_B\approx
\frac{1}{2\pi}\frac{e^2\sqrt{\pi}}{L^2}\frac{F}{V_{T,f}}
\label{eq:harmonicBund}
\end{equation}
In Fig.~\ref{fig:harmonicError}(c) we plot the value of
$V_{T,f}/F$ needed to observe $N_{B}$ Bloch oscillations for
lattice sizes $L=64$ and $100$; note that it is a $log$-$log$
plot. The filled circles represent the point for 50000 BO and the
arrows below indicate the trap strengths required. For the
experimentally relevant lattice sizes between $L=50$ and $100$,
the trap strength needs to be extremely small, e.g., $V_{T,f}/F
\approx 10^{-8}$, to observe BO cycles beyond the current maximum
value of 20000~\cite{nagerl10}. Larger lattice sizes make the
constraint more severe.

If the initial pre-quench state is trapped, it causes density
inhomogeneity and $N_{B}$ increases up to 20$\%$. Then $N_{B}$
depends on a combination of initial trap, total atom number and
the density profile, and specific cases must be analyzed
numerically. We note that the momentum width for an initially
trapped case is bigger, and it spreads more quickly during the
dynamics, eventually making the number of observable BO smaller.
The overall effect of an initial trap is not significant compared
to that of the residual trap, and hence our analysis here using an
initially homogeneous density profile gives a good approximation
for the bounds on $V_{T,f}$.

\begin{figure}[ptb]
\vspace{-0.2cm}
\par
\begin{center}
\includegraphics[width=0.35\textwidth,angle=0]{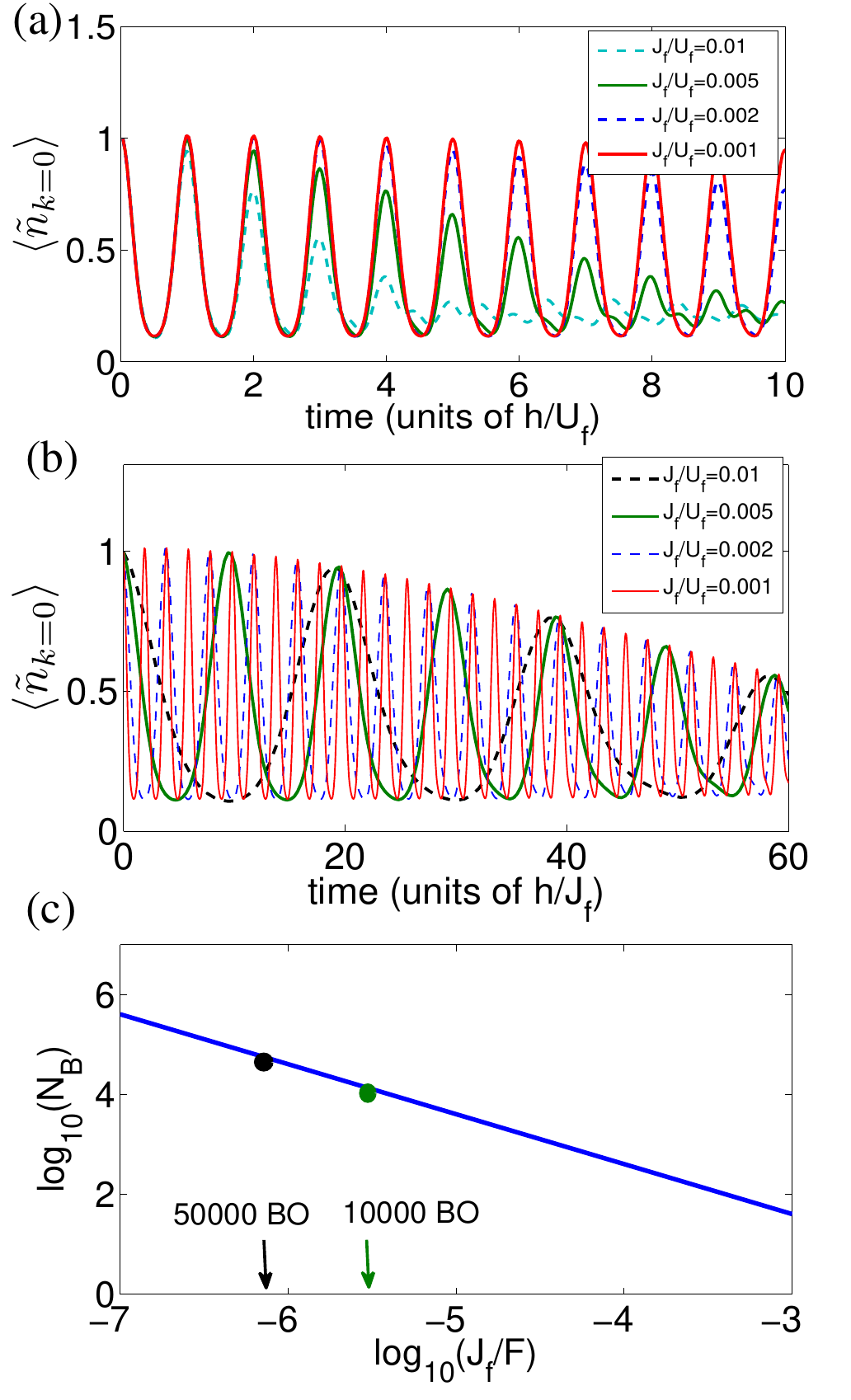}
\end{center}
\par
\vspace{-0.6cm}\caption{(color online) Bounds on the number of
Bloch oscillations due to finite-$J_f$. Panel (a) depicts the
$\langle \tilde{n}_{k=0} \rangle$ collapse and revival signal
decay for different values of $J_f/U_f$, as a function of hold
time, in units of interaction time-scale $h/U_f$. The homogeneous
initial superfluid state has $U_{i}/J_{i}=0$ and $\bar{n}=1.5$. As
$J_f/U_f$ increases, the number of observable oscillations
decrease as expected. The revivals occur approximately at integer
multiples of $h/U_f$, deviating slightly for increasing $J_f/U_f$.
In (b) we plot the same data as in (a), but in units of tunneling
time $h/J_f$. We see that the damping of the signals can be
described by a common envelope function. Finding the decay for one
value of $J_f/U_f$, we can find the damping of $M$th-oscillation
by a simple scaling. Panel (c) shows the number of observable
Bloch oscillations $N_B$ for different values of $J_f/F$; to
observe 50000 BO, $J_f/F \approx 0.8 \times 10^{-6}$.}
\label{fig:Jerror}
\end{figure}

\subsection{Bounds on finite-J effects}

In an ideal BO collapse and revival scenario $J_f=0$ and the
momentum peak revives completely. For $J_f \neq 0$,
Fig.~\ref{fig:finiteJ} showed that CR oscillation amplitudes
slowly decay, and similarly the BO signal dephases due to the
competition among $J_f, U_f$ and $F$. Here we analyze the bounds
on the number of Bloch oscillations for a finite non-zero value of
$J_f/U_f$ and $J_f/F$. For this, we assume that $J_f$ is small: $F
\gg J_f$ and $U_f \gg J_f$.

In Fig.~\ref{fig:Jerror} we analyze the decay of CR oscillations
of $\langle \tilde{n}_{k=0} \rangle$ for different values of
$J_f/U_f$ and $F=0, V_{T,f}=0$, using TEBD numerical simulations.
The initial superfluid state is a homogeneous coherent state with
$U_i/J_i=0$, $\bar{n} \approx 1.5$ and $L=32$.
Figure~\ref{fig:Jerror}(a) shows the dynamics in units of
interaction timescale $h/U_f$ where the expected higher rate of
decay for larger tunneling values is evident. If the same data is
plotted in units of $h/J_f$, as shown in Panel (b), a common
envelope function is seen to characterize the decay of $\langle
\tilde{n}_{k=0} \rangle$. This implies that for a specific value
of tunneling $J_f/U_f$, the signal decay after one oscillation is
equal to that of the $M$th oscillation for the smaller value
$\frac{1}{M}J_f/U_f$. The reference $J_f/U_f$ needs to be small
for this relationship to hold. For values considered in
Fig.~\ref{fig:Jerror}(b), this holds true. The damping of the
first oscillation revival analyzed in
Refs.~\cite{fischer08,rigol10} is consistent with our findings. We
show here that this analysis can be extended to the $M$th
oscillation, and propose a method to estimate the number of
observable oscillations for smaller tunneling rate by calculating
the oscillation decay for a larger one.

We can also make a connection with the number of observable BO,
again defined by the time at which the envelope of $\langle
\tilde{n}_{k=0} \rangle$ reaches $1/e$ of its initial value. In
the presence of finite-J, the CR oscillations due to $U_f$ and BO
due to $F$ are coupled as discussed in Sec. V.B. In the regime of
interest, when $F \gg J_f$ and $U_f \gg J_f$, the effects of $U_f$
and $F$ on the oscillations are approximately separable.
Figure~\ref{fig:Jerror}(c) depicts the Bloch oscillations $N_{B}$
that can be observed for different values of $J_f/F$. To observe
50000 BO cycles, $J_f/F$ value needs to be $0.8 \times 10^{-6}$.

The value of $J_f$ depends on optical lattice depth $V$ in the
following way: $\frac{J_f}{U_f}=\frac{2\sqrt{2}d}{\pi a_s} e^{-2
\sqrt{V}}$, where $V$ is given in units of recoil energy $E_r$,
$a_s$ is the s-wave scattering length, and $d$ is lattice spacing.
We estimate that quenching to $V > 20 E_r$ puts us into a regime
of $J_f/U_f > 10^{-6}$ and depending on the ratio of $F$ and
$U_f$, $J_f/F > 10^{-6}$ can be achieved.

The damping of BO due to finite-$J$ depends on several things: the
initial average occupation, $J_i/U_i$, initial trap $V_{T,i}$ and
force $F$, in addition to its dependence on $J_f/U_f$. We find
here that knowing all the other parameters, the bounds on $J_f/F$
and $J_f/U_f$ to observe $N_B$ oscillations has a linear
dependence. We have not discussed even longer term behavior of the
decay as the question of thermalization and equilibration can
become important~\cite{kinoshita06,rigol08}. The value of
$J_f/U_f$ should be small in precision measurement experiments
such that a large number of Bloch oscillations can be observed
before equilibration takes place.

\section{Conclusion and Summary}
\label{sec:conclusion}

In this paper, we have shown that the effect of multi-particle
interactions on Bloch oscillations physics is described by the
physics of matter-wave revivals -- full revivals for a decoupled
lattice and partial revivals for a coupled lattice, all occurring
on the interaction timescale. We performed a theoretical analysis
of interacting ultracold bosons in a suddenly ramped
one-dimensional optical lattice that is vertically aligned. This
set up can be systematically tuned and exploited to study the
effects of interactions on BO. We used the Bose-Hubbard
Hamiltonian to model the dynamics in the strongly interacting
regime, and studied the dynamics in two limits -- the strong-$U$
($U>F$) regime, and the strong-$F$ ($F>U$) regime, where $U$ and
$F$ are respectively atom-atom interaction and linear potential
strength, after the quench. We have used the time-evolving block
decimation (TEBD) algorithm for our numerical simulations.

We analyzed three dephasing mechanisms for the oscillations --
effective three-body interactions, finite value of tunneling $J$
and residual harmonic trapping. We find that the dephasing effect
due to effective three-body interactions becomes important for the
strong-$U$ regime. When $J\neq0$, we predict that Bloch
oscillations of the center of mass of the atomic cloud should also
go through collapse and revival modulations, demonstrating an
example of quantum transport where real-space revivals occur. We
also show that the presence of a harmonic trap during the dynamics
quickly destroys coherence visibility of the atoms and gives rise
to a temporal Talbot effect~\cite{chapman95,nagerl10}, which
survives in the strongly interacting few-atom regime. We further
model in detail the momentum and real space oscillations of a
lattice-trapped superfluid in the presence of gravity, a residual
harmonic potential and finite tunneling.

In addition to studying the interplay between interactions and
Bloch oscillations physics, we examine the prospects of measuring
the gravitational acceleration $g$ with high precision using a
system of strongly-correlated ultracold atoms in a deep lattice.
We present numerical and analytical results for error bounds on
the residual harmonic trap and finite tunneling to go beyond the
current maximum observation of 20000 Bloch oscillations. The
analysis and characterization of a realistic experimental setup is
a necessary step towards the goal of surpassing the current
precision limit of $g$.

The ideas investigated here are extremely relevant in the light of
current experimental efforts~\cite{nagerl13}. Further insights can
be gained by studying more comprehensively the competition of $F$,
$U$ and $J$. Realistic conditions such as finite temperature and
higher-band effects may also be relevant for cold atom
experiments. The effects of interactions on Bloch oscillations in
cold atoms and other systems deserve additional analysis for the
exploration of fundamental physics as well as measurement
applications.

\begin{acknowledgments}
We acknowledge support from the US Army Research Office under
Contract No. 60661PH and the National Science Foundation Physics
Frontier Center located at the Joint Quantum Institute.
\end{acknowledgments}

\end{document}